\documentclass[twocolumn,aps,showpacs,epsfig]{revtex4}

\usepackage{amsmath}
\usepackage[dvips]{graphics}
\usepackage{latexsym}
\usepackage{subfigure}

\DeclareFontFamily{OT1}{rsfs}{}
\DeclareFontShape{OT1}{rsfs}{m}{n}{ <-7> rsfs5 <7-10> rsfs7 <10->
rsfs10}{} \DeclareMathAlphabet{\mycal}{OT1}{rsfs}{m}{n}
\def\scri{{\mycal I}}

\begin{document}
\newcommand{\bea}{\begin{eqnarray*}}
\newcommand{\eea}{\end{eqnarray*}}
\newcommand{\bean}{\begin{eqnarray}}
\newcommand{\eean}{\end{eqnarray}}
\newcommand{\eqs}[1]{Eqs. (\ref{#1})} 
\newcommand{\eq}[1]{Eq. (\ref{#1})} 
\newcommand{\meq}[1]{(\ref{#1})} 
\newcommand{\fig}[1]{Fig. \ref{#1}}

\newcommand{\tri}{\delta}
\newcommand{\grad}{\nabla}
\newcommand{\pa}{\partial}
\newcommand{\pf}[2]{\frac{\pa #1}{\pa #2}}
\newcommand{\cla}{{\cal A}}
\newcommand{\aqt}{\frac{1}{4}\theta}

\newcommand{\om}{\omega}
\newcommand{\omo}{\omega_0}
\newcommand{\ep}{\epsilon}
\newcommand{\nonu}{\nonumber}
\newcommand{\scrip}{\scri^{+}}
\newcommand{\hp}{{\cal H^+}}
\newcommand{\tm}{\tilde M} 
\newcommand{\ts}{\frac{\sqrt 3}{2}}
\newcommand\tabcaption{\def\@captype{table}\caption}

\title{Local conditions for the generalized covariant entropy bound}

\author{Sijie Gao}
\email{sijie@fisica.ist.utl.pt}
\author{Jos\'e P. S. Lemos}
\email{lemos@fisica.ist.utl.pt}
\affiliation{
Centro Multidisciplinar de Astrof\'{\i}sica - CENTRA,\\
Departamento de F\'{\i}sica, Instituto Superior T\'ecnico,\\
Universidade T\'ecnica de Lisboa,\\
Av. Rovisco Pais 1, 1049-001 Lisboa, Portugal}

\begin{abstract}
A set of sufficient
conditions for the generalized covariant entropy bound given by
Strominger and Thompson is as follows: Suppose that the entropy of 
matter can be described by an entropy current $s^a$. Let $k^a$ be any
null vector along $L$ and $s\equiv -k^a s_a$. Then the generalized
bound can be derived from the following conditions: (i) $s'\leq 2\pi
T_{ab}k^ak^b$, where $s'=k^a\grad_a s$ 
and $T_{ab}$ is the stress energy tensor; (ii) on the initial
2-surface $B$, $s(0)\leq -\frac{1}{4}\theta(0)$, where $\theta$ is
the expansion of $k^a$. We prove that condition (ii) alone can be used
to divide a spacetime into two regions: The generalized entropy
bound holds for all light sheets residing in the region where
$s<-\frac{1}{4}\theta$ and fails for those in the region where
$s>-\frac{1}{4}\theta$. We check the validity of these
conditions in FRW flat universe and a scalar field spacetime. Some
apparent violations of the entropy bounds in the two spacetimes 
are discussed. These holographic bounds are important in the formulation 
of the holographic principle. 
\end{abstract}

\pacs{04.70.Dy, 04.60.-q}

\maketitle

\section{Introduction}
Bounds on entropy set by some specified area that surrounds a certain 
volume are called holographic bounds and are important in the 
formulation of the holographic principle of 't Hooft. There are several 
such bounds, the one that concerns us here is a generalization of 
the covariant entropy bound. 
The covariant entropy bound, conjectured by Bousso, 
is the following
\cite{bousso,bn,three}: Let $B$ be a spacelike 2-surface in a
spacetime $(M,g_{ab})$ satisfying Einstein's equation and the dominant
energy condition. Its area is denoted by $A_B$.  Consider a null
hypersurface $L$ generated by null geodesics, each with tangent
vector field $k^a$ which starts at $B$ and is orthogonal to
$B$. Suppose that the expansion
\bean
\theta=\grad_a k^a \label{epn}
\eean
of $k^a$ is non-positive everywhere on $L$ and $L$ is not terminated until a
caustic is reached (where $\theta\rightarrow -\infty$). Then the
entropy, $S_L$, through $L$ satisfies 
\bean
S_L\leq \frac14\, A_B\,. \label{sbd}
\eean
Evidences supporting this bound have been studied in situations where
other non-covariant bounds fail, such as in cosmological spacetimes
and other matter systems \cite{bousso,bn,three,viqar,gaolemos}. The
null surface $L$ in the conjecture is required to be extended as far
as possible unless a caustic is reached. Flanagan {\em et. al}\,
\cite{three} modify this bound by allowing $L$ to be terminated at
some spacelike 2-surface $B'$ before coming to a caustic. Then the
inequality \meq{sbd} is replaced by 
\bean
S_L\leq \frac{1}{4}\,(A_B-A_{B'}). \label{get}
\eean 
This is called the generalized covariant entropy bound, or 
the generalized Bousso bound.

Recently, Strominger and Thompson \cite{quan} suggested a set of
simple assumptions from which the generalized bound \meq{get} can
be derived. The authors 
in \cite{quan}  assumed that the matter entropy can be
described in terms of an entropy current $s^a$, where $s^a$ is
independent of the null surface $L$.  Apart from this, the following
two conditions are postulated: 
\vskip 0.1cm
\noindent
(i) Let 
\bean
s\equiv -k^a s_a \,\label{firstcondition_a}
\eean 
be the flux of entropy that crosses the light sheet $L$. Let  
\bean
s'=k^a\grad_a s \,\label{firstcondition_b}
\eean 
be the rate of entropy flux on $L$ and $T_{ab}$ be the stress-energy tensor. 
Then 
\bean
s'\leq 2\pi T_{ab}k^a k^b\,, \label{co}
\eean
i.e., the rate of entropy flux is less than the energy flux 
throughout the  light sheet. 

\vskip 0.1cm\noindent
(ii) On the initial 2-surface $B$, where the affine parameter
$\lambda$ is set to zero,  
\bean
s(0)\leq-\frac{1}{4}\,\theta(0). \label{ct}
\eean

Only local quantities are
involved in conditions (i) and (ii). Bousso, {\em et. al} 
\cite{simple} suggested a similar set of sufficient conditions 
which is stronger, and we do not discuss it here. 

Conditions \meq{co} and \meq{ct}  apply to spacetimes where absolute
entropy currents 
(i.e., entropy currents that do not depend on the light sheet) are
well defined. Moreover, as we shall show explicitly in Proposition
\ref{prop}, condition (i) becomes superfluous for testing the 
generalized bound, when condition (ii) is regarded as a pointwise
condition, in which case it gives a straightforward criteria for
the generalized covariant bound. Thus, we suggest an even simpler 
assumption from which the generalized entropy bound can be derived:
\vskip 0.1cm
\noindent 
(A) Given a spacetime region one has 
\bean
s(x)\leq-\frac{1}{4}\,\theta(x)\,, \label{glcondition}
\eean
where $x$ represents any spacetime point within the region. 

This is 
just a generalization of Thomson and Strominger's condition (ii), from 
an initial surface, to the whole region. Armed with this condition 
we can prove Proposition \ref{prop} (see next section for the 
precise formulation and proof), 
which states 
that a  necessary and sufficient  condition 
for the generalized entropy bound 
to be satisfied for all light sheets in a region, is that
condition (A) (i.e., $s\leq-\frac{1}{4}\theta$)  
is satisfied. 
Then, the bound holds in the region where
$s<-\frac{1}{4}\theta$ and fails in the region where
$s>-\frac{1}{4}\theta$. 
There are, however, some violations. Indeed, since $\theta\simeq0$ for
light sheets in a small neighborhood of the apparent horizon, this
result indicates that if the apparent horizon is located in a matter
system (where $s=s_{\rm matter}>0$) the generalized covariant entropy
bound is generally violated for light-sheets in a sufficiently small
neighborhood of the apparent horizon. In such a neighborhood one
surely has $s>-\frac{1}{4}\theta$.  However, as we will see along the
paper, such kind of violation is due to a collapse of the hydrodynamic
description of matter entropy for light sheets in a small neighborhood
of the apparent horizon.  We therefore treat this kind of violations
as trivial violations (see also Bousso {\em et. al} \cite{simple}).

We shall apply Proposition \ref{prop} to two cases, namely, to
a closed Friedmann-Robertson-Walker (FRW) universe and to
a scalar field spacetime. In both cases absolute entropy currents
$s^a$ can be defined.

In the first case, a FRW universe, we test condition (A) and show that
it is valid throughout the spacetime except for regions very close to
the singularity and the apparent horizon. Thus, following 
the previous discussion, we conclude that there
is no meaningful violation to the generalized covariant entropy bound
in the cosmological spacetime. Then from Proposition \ref{prop} we
know that the generalized covariant entropy bound holds. For
completeness we also test Thomson and Strominger conditions (i) and
(ii) (although (ii) is automatically satisfied when (A) is satisfied)
and show they hold. In this manner we complete the analysis made in
\cite{bousso} for some selected light sheets in a FRW universe.

In the second case, a time dependent spherically symmetric massless scalar 
field spacetime, the issue is more interesting, since it yields a new example
for testing the bounds.  This spacetime is
characterized by a past spacelike singularity, a timelike naked
singularity and an apparent horizon.  Such a study has been initiated
by Husain \cite{viqar} where covariant entropy bounds in such a time
dependent spherically symmetric massless scalar field spacetime were
examined.  We reexamine this issue by checking the sufficient
condition (A) (see Equation (\ref{glcondition})) and make improvements
in the following aspects: 
First, we find that the formula of the entropy density proposed in
\cite{viqar} is valid only in a region that does not include the
neighborhood of the naked timelike singularity. 
Second, Husain \cite{viqar} identified, in an unusual
energy-temperature relation, the coefficient $\sigma$ 
with the Stefan-Boltzmann constant. However, $\sigma$ remains
undetermined and the value $\sigma=\frac{1}{2}$ adopted in
\cite{viqar} is by no means generic. Using $\sigma=\frac12$, 
Husain found no violation for the original Bousso bound
(\ref{sbd}). On the other hand we find that in some regions of this
spacetime the bound is violated. We check the Bousso bound for light
sheets hitting the past spacelike singularity and find that the Bousso
bound is violated for such light sheets starting near the apparent
horizon. This violation can not be explained by the failure of the
fluid description for short light sheets, but can be rescued by 
a smaller value of $\sigma$. Thus, 
by using the covariant entropy bound one can put an upper limit in
$\sigma$.
Third, Husain \cite{viqar} found that the generalized bound is
violated for light sheets in the neighborhood of the apparent horizon.
This can be easily explained by our Proposition \ref{prop}, since as
we have argued, near the apparent horizon $\theta\simeq0$, and it is
trivial to have matter satisfying $s>-\aqt$ in violation of the
bound. By calculating $s(x)$ and $\theta(x)$ we can find the boundary
hypersurface $s=-\frac14\theta$. According to Proposition \ref{prop}
the bound is violated in the region in-between the apparent horizon and 
this boundary.
Fourth and final, a length scale argument \cite{simple} has been 
used to eliminate
counterexamples for the generalized bound. 
This argument states that if the
proper distance of the path traveled by the light sheet is smaller
than the thermal wavelength of the matter then the hydrodynamic
description for the matter fails.  Now, Husain \cite{viqar} showed
that for some light sheets the bound is violated and this
violation cannot be explained by the length scale argument. Thus, in
these cases, the argument is inconclusive to eliminate
counterexamples for the generalized bound. In our analysis, we reach a
similar conclusion. However, the length scale used in \cite{viqar} is
a black body length scale, $\rho^{-1/4}$, which is not consistent with
our analysis of the scalar field, where we show one should use an
associated length scale of the form $\rho^{-1/6}$.

In this paper, we use units with $c=k_B=\hbar=G=1$ and the $(-,+,+,+)$
metric signature.

\section{Proposition} \label{proof} 
In this section, we propose a simple but useful criteria for the 
generalized covariant entropy bound. 
\newtheorem{prop}{Proposition}
\begin{prop} \label{prop}
A necessary and sufficient  condition 
for the generalized entropy bound to
be satisfied for all light sheets in a region is that
condition (A), i.e., $s\leq-\frac{1}{4}\,\theta$, 
is satisfied everywhere in the region.
\end{prop}
{\bf Proof}:  We use the coordinate system
$(\lambda,x^1,x^2)$ introduced in \cite{three} to describe the light
sheet $L$, where $(x^1,x^2)$ is any coordinate system on the initial
two-surface $B$ and
$\lambda$ is the affine parameter of the the null generators. Associated with 
each generator, one can define an area-decreasing factor \cite{three}:
\bean
{\cal A}(\lambda)\equiv \exp\left[\int_0^\lambda d\bar\lambda \theta(\bar
\lambda)\right], \label{adf}
\eean 
which has the following obvious properties:
\bean
{\cal A}(0)=1, \label{pone}
\eean
and
\bean
{\cal A}'(\lambda)=\cla(\lambda)\theta(\lambda). \label{ptwo}
\eean
Then the entropy crossing $L$ can be expressed as \cite{three} 
\bean
S_L=\int_Bd^2x\,\sqrt{\det
  h_{AB}(x)}\int_0^{\lambda_\infty(x)}d\lambda\,
s(x,\lambda)\,\cla(x,\lambda), \label{imfs}
\eean 
where $x\equiv(x^1,x^2)$ is a point on $B$,  
$h_{AB}(x)$ is the induced
2-metric on $B$, $\lambda_\infty(x)$ is the affine parameter at the
endpoint of the generator which starts at $x$ on $B$ and ends on
$B'$, $s(x,\lambda)$ is the entropy density at the point $(x_1,x_2,\lambda)$, 
and $\cla(x,\lambda)$ is the area-decreasing factor for 
the null generator that starts at point $x$. 
Following \cite{three}, we shall rescale the affine parameter along
each generator such that $\lambda_\infty (x)=1$ at $B'$.
Then, from \eqs{ptwo} and \meq{imfs},  
if $s\leq-\frac{1}{4}\,\theta$ in the neighborhood of the light sheet, we
have 
\bean
S_L&\leq& -\frac{1}{4}\int_Bd^2x\sqrt{\det
  h_{AB}(x)}\int_0^{1}d\lambda\,
\cla'(x,\lambda) \nonumber  \\
&=&\frac{1}{4}\int_Bd^2x\sqrt{\det h_{AB}(x)}\,[\cla(x,0)-\cla
(x,1)] \nonumber \\
&=&\frac{1}{4}\, (A_B-A_B'). \label{saa}
\eean
In the last step, we have used the formulas given in \cite{three}:
\bean
A_B&=&\int_B d^2x\sqrt{\det h_{AB}(x)}\,,  \label{abfa} \\
A_{B'}&=&\int_B d^2x\sqrt{\det  h_{AB}(x)}\cla(x,1)\,.
 \label{abf}
\eean
Note that the result in \eq{saa} is independent of the rescaling 
for $\lambda_\infty(x)$. Thus, we have proved the sufficient part. 

Now we prove the necessary part of the 
proposition. If the inequality \meq{saa} holds for all light sheets  
in a region, choose any one of them, and for convenience do 
a different rescaling of  the affine parameter, such 
that $\lambda_\infty(x)=\Delta\lambda$. 
Let $B$ shrink to a
sufficiently small area and let $B'$ be sufficiently close to
$B$ (i.e., $\Delta\lambda \rightarrow 0$). Then we obtain the ``local
version'' of \eq{imfs} 
\bean
S_L=s(0)\,A_B\,\Delta\lambda, \label{ifs}
\eean
where $s(0)$ is the entropy density on $B$ (since $B$ has shrunk to
an arbitrarily small area, $s$ can be treated as a constant on $B$)  and
we have used \eqs{pone} and \meq{abfa}. 
On the other hand, \eq{abf} becomes
\bean
A_{B'}&=&\int_B d^2x\sqrt{\det
  h_{AB}(x)}\cla(\Delta\lambda) \nonumber \\
&=& A_B[\cla(0)+ \cla'(0)\Delta\lambda 
]. 
\eean
Hence,
\bean
A_{B}-A_{B'}=-A_B\,\cla'(0)\, \Delta\lambda. \label{abbp}
\eean
Substituting \eq{ifs} and \eq{abbp} into  \eq{get} we get condition (A), 
which is what we wanted to show.

By investigating condition (A), we shall be able to 
identify the regions where the bound holds for all the light sheets lying
inside it. It is worth noticing that Proposition \ref{prop} does not
cover the case when a light 
sheet crosses both the $s>-\frac14\,\theta$ zone and the 
$s<-\frac14\,\theta$ zone. 

Our next remark explores the implication of Strominger and Thompson's 
sufficient conditions (i) and (ii). 

\medskip
\noindent {\bf Remark}\;\; \label{exte}
{\it Suppose conditions (i) and (ii) in \cite{quan} are satisfied for a light
sheet $L$. Then $s\leq-\frac{1}{4}\,\theta$ is satisfied at all 
points on $L$. }
\medskip

The proof of this remark was almost given in  \cite{quan}. 
Under the two conditions, it is shown in \cite{quan} that 
\bean
s(\lambda)\leq-\frac{1}{2}\frac{G'(\lambda)}{G(\lambda)}, \label{squ}
\eean
where $G\equiv \sqrt{\cla}$. It then follows immediately, from \eq{ptwo}, that 
$s\leq-\frac{1}{4}\,\theta$ all over the light sheet. 

Proposition 1 and the Remark imply that the role of condition (i) could almost 
be replaced by the single condition (A). If condition (A)
is satisfied everywhere on the light sheet, then Proposition \ref{prop}  
guarantees that the generalized Bousso bound is satisfied, no matter whether 
condition (i) is violated or not. On the other hand, if condition (A) holds
on the initial surface $B$ but not throughout the light sheet, the Remark 
means that condition (i) must fail at some points on the light 
sheet. So verifying condition (i) will not help judge the generalized Bousso
bound. In section \ref{subfi}, we show that there exist light sheets
which satisfy the generalized Bousso bound, but condition (i) fails
everywhere on the light sheets. Therefore, condition (i) is not a
necessary condition for the bound. Further implications of these
conditions will be discussed in the applications of the following section.

\section{Applications}

\subsection{Application to a closed universe}
We consider a closed, dust-dominated FRW universe which is described
by the metric
\bean
ds^2=
a^2(\eta)(-d\eta^2+d\chi^2+\sin^2\chi d\theta^2+\sin^2\chi\sin^2\theta
d \varphi^2)\,, \nonumber \\
&& \label{frwm}
\eean  
where 
\bean
a(\eta)=\frac{a_{\rm max}}{2}(1-\cos\eta)\,. \label{aet}
\eean
In order to test our condition (A) and condition (i) of 
Thompson and Strominger \cite{quan} we need some preliminaries.
The Planck proper time, $\tau_{PL}$ (where 
$\tau_{PL}\sim1$), corresponds to the
coordinate time $\eta=\eta_{PL}$, which is obtained from
\eqs{frwm} and \meq{aet} as 
$\eta_{PL}\sim a_{\rm max}^{-1/3}$ \cite{bousso}. 
Now we choose the future-directed
outgoing light sheet such that its tangent vector is given by 
\bean
k^a=(\pa/\pa\eta)^a+(\pa/\pa\chi)^a. \label{koa}
\eean
Since such a universe is homogeneous and evolves adiabatically, the
physical entropy density ${s_{}}_{\rm phys}$ 
takes the form \cite{earlyuni} 
\bean
{s_{}}_{\rm phys}(\eta)=\frac{s_0}{a^3(\eta)},  \label{cen}
\eean
where $s_0$ is a constant. $s_0$ can be specified by the requirement
that the entropy density may not exceed one at the Plank time. 
Since, with the help of \eq{aet}, $a^3(\eta_{PL})\sim
a_{\rm max}$, we have $s_0\sim a_{\rm max}$.
Note also that the four-velocity of a comoving observer is 
\bean
u^a=\frac{1}{a(\eta)}\left(\frac{\pa}{\pa \eta}\right)^a\,. \label{fvc}
\eean
Thus, the entropy flux can be constructed as
\bean
s^a={s_{}}_{\rm phys}(\eta)\,u^a
=\frac{s_0}{a^4(\eta)}\left(\frac{\pa}{\pa \eta}\right)^a.
\label{ef}
\eean
Then the entropy associated with $k^a$ is 
\bean
s(\eta)=-s_a k^a=\frac{4}{a_{\rm max}(1-\cos(\eta))^2}. \label{sk}
\eean
To test condition (A), we calculate the expansion of $k^a$,
\bean
\theta=2 \left(\cot(\eta/2)+\cot\chi \right). \label{thek}
\eean
It is easy to see that testing condition (A) 
is equivalent to testing the following inequality
\bean
\left|\frac{4 s}{\theta}\right|=\left|\frac{\csc^3\eta\,
  \csc(\eta/2+\chi)\,\sin\chi}{a_{\rm max}}\right|\leq 1. \label{inet}
\eean
In Plank units, $a_{\rm max}$ is a very large number. Then very close
to the singularity, $\eta<\eta_{PL}$, one has that the expression inside
the absolute value in \eq{inet} goes as $|\eta_{PL}^3/\eta^3|$, and so
the inequality \eq{inet} fails in that region.  This is not a real
violation of the bound because the region is inside the quantum
regime.  In addition, the inequality may fail around
$\chi=\pi-\eta/2$, 
where $\theta=0$. Now, this is where the apparent horizon is
located. Thus, as long as a light sheet does not go sufficiently close
to the apparent horizon (by definition, a light sheet can never cross
the apparent horizon), the generalized entropy bound always holds. As
discussed in the introduction, light sheets that are located in a
small neighborhood of the apparent horizon should not be of our
concern since the hydrodynamic description fails there.  We conclude
that the generalized entropy bound always holds if light sheets are
reasonably chosen (not too close to the singularities or to the
apparent horizon).

Now we test condition (i) of Thompson and Strominger \cite{quan}. 
The derivative of the entropy flux, $s'$,  in condition (i) 
for the FRW universe (see \eq{frwm}) is 
given by  
\bean
s'(\eta)=k^a\grad_a s=-\frac{8 \sin\eta}{a_{\rm max}(1-\cos\eta)^3}.
  \label{skp} 
\eean
It is also straightforward to compute the energy flux through 
the light sheet, 
\bean
T_{ab}k^a k^b =\frac{3}{4\pi(1-\cos\eta)}. \label{tkk}
\eean 
Since $T_{ab}k^a k^b$ is positive, to test condition (i), it is
sufficient to test the following inequality
\bean
\frac{|s'(\eta)|}{2\pi T_{ab}k^a k^b}=\frac{16|\sin\eta|}{3
  a_{\rm max}(1-\cos\eta )^2} \leq 1. \label{ra}
\eean
In Plank units, $ a_{\rm max}$ is a very large number. So inequality \meq{ra}
can be violated only when $\eta$ is sufficiently small (or by
symmetry, sufficiently close to $2\pi$). The detail can be seen by
power series expansion of $\frac{|s'(\eta)|}{2\pi T_{ab}k^a k^b}$ 
around $\eta=0$. The leading term is
\bean
\frac{|s'(\eta)|}{2\pi T_{ab}k^a k^b}
\sim \frac{64}{3a_{\rm max}\eta^3}. \label{ter}
\eean
Therefore, only when 
$\eta \lesssim 2.77 \, a_{\rm max}^{-1/3}=2.77 \eta_{PL}$,
can inequality \meq{ra} be violated. However, this violation takes
place within a few  Plank times from the singularity, where quantum gravity
takes effect. Thus, the violation does not occur in the classical regime. 
Condition (ii) of Thompson and Strominger \cite{quan} is a particular 
instance of our condition (A) and therefore has been tested above.

\subsection{Application to a scalar field spacetime}
\subsubsection{The exact scalar field solution}

We shall now review the scalar field solution presented in
\cite{viqar,ss}. From the Einstein-scalar field equations
for massless minimally coupled scalars, a spherically symmetric
solution is given by the metric
\bean
ds^2&=&-t\,f(r)\,dt^2+t\,f(r)^{-1}\,dr^2 \nonumber \\
&&+t\,r^2 \,f(r)^{-\frac{1+\sqrt 3 /2}{\sqrt 3/2}}
\left( d\theta^2+  \sin^2\theta d\phi^2\right), \label{st}
\eean 
where 
\bean
f(r)=\left(1-\frac2r\right)^{-\frac{\sqrt3}{2}}\,.
\label{fr}
\eean
(There is also a possibility of choosing $-\frac{\sqrt{3}}{2}$ where
there is  $\frac{\sqrt{3}}{2}$ in \eqs{st} and \meq{fr} but we do not
consider it here).  
The corresponding scalar field in the spacetime is 
\bean
\phi(t,r)=\frac{1}{4\sqrt \pi}\ln\left[t^{\sqrt3}\,
f(r)^{1/\sqrt3}\right].  \label{phi}
\eean

For future discussions, we introduce the following properties of the
spacetime. The  future directed 
tangent vector field of the radial ingoing null geodesics is 
\bean
k^a=\frac{1}{f(r)\,t}\left(\pf{}{t}\right)^a-\frac{1}{t}
\left(\pf{}{r}\right)^a\,.
\label{ig} 
\eean
The corresponding expansion is 
\bean
\theta=\frac{1}{t}-\frac{1}{t_{\rm AH}(r)}, \label{ex}
\eean
where
\bean
t_{\rm
  AH}(r)=\frac{r^2}{2\left(r-1+\sqrt{3}/2\right)}\,f(r)^{-1-\frac{2}{\sqrt 3}},
  \label{tah}  
\eean
and the equation $t=t_{\rm AH}(r)$ yields the apparent
horizon as a function of $r$. 

\begin{figure}[htpb]
\centering
\scalebox{0.8}
{\includegraphics{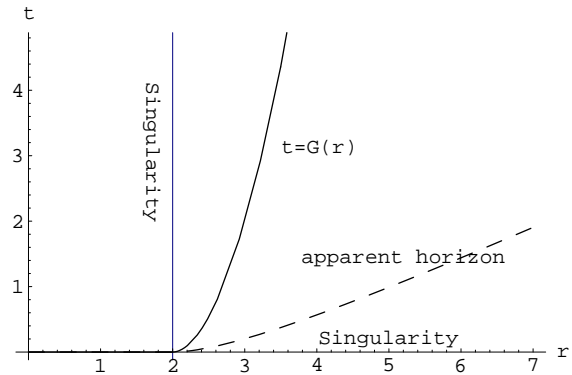}}
\caption{A spacetime diagram for the scalar field spacetime.  Below the
solid line $t=G(r)$, $\pa^a\phi $ is timelike. Between $t=G(r)$ and
the timelike singularity $r=2$, no observers associated with the
scalar field can be naturally defined.
}  
\label{timapp}
\end{figure}

In order to proceed our investigation in a perfect fluid context,
we need to define comoving observers. As suggested in
\cite{viqar}, one may choose the four-velocity of the observers to be
parallel to   $\partial^a\phi$. This requires that $\partial^a\phi$ be
timelike.  Using \eq{st} one finds
\bean
\partial_a\phi=\sqrt{\frac{3}{\pi}}\,\frac{1}{4t}\,
dt_a+\frac{f'(r)}{4\sqrt{3\pi}f(r)}\,dr_a \,.\label{dp} 
\eean
Then a straightforward calculation
shows  that $(\pa\phi)^2\equiv\pa_a\phi\,\pa^a\phi$,  given by 
\bean
&&(\partial\phi)^2=g^{ab}\pa_a\phi\pa_b\phi \\
&=& \frac{1}{48\pi
  t^3}\left(\frac{r-2}{r}\right)^{\sqrt 3/2}\left(-9+
\frac{3\left(\frac
  {r-2}{r}\right)^{-\sqrt 3} t^2}{(r-2)^2r^2}\right) \label{pfs}
\eean
is negative only for 
\bean
t<G(r)\equiv \sqrt{3} (r-2)^{1+\sqrt{3}/2}r^{1-\sqrt{3}/2}\,. \label{ttr} 
\eean
Thus, our discussion should be confined to the region $t<G(r)$, where
$\partial^a \phi$ is timelike and
comoving observers can exist. As mentioned in \cite{viqar}, the
spacetime has two curvature 
singularities located at $t=0$ and $r=2$.   Fig. \ref{timapp} shows the
plots of the two singularities, the apparent horizon, and 
$t=G(r)$.  

Now we review the derivation of the entropy flux 4-vector in the
framework of \cite{viqar}. Assume the scalar field is a perfect fluid
with stress-energy tensor
\bean
T_{ab}=\rho u_a u_b+P( u_a u_b+g_{ab}). \label{trp}
\eean
On the other hand, the stress-energy tensor can be calculated from the
metric \meq{st} as
\bean
T_{ab}=\pa_a\phi\,\pa_b\phi-\frac{1}{2}g_{ab}(\pa\phi)^2\,. \label{tm}
\eean
Define 
\bean
u^a=-\frac{\pa^a\phi}{\sqrt{-(\pa\phi)^2}} \label{fov}
\eean
and rewrite \eq{tm} as 
\bean
T_{ab}=-\frac{1}{2}(\pa\phi)^2 u_a u_b-\frac{1}{2}(\pa\phi)^2(u_a
u_b+g_{ab}). \label{rst}
\eean
When $\pa^a\phi$ is timelike, $u^a$ may be identified as the 4-velocity
of observers comoving with the scalar field. 
Therefore, by comparing \eq{trp} and \eq{rst}, we have
the following relation
\bean
P=\rho=-\frac{1}{2}(\partial\phi)^2. \label{per}
\eean

Now define 
\bean
\omega=k \,,\label{omk}
\eean
where  $k$ is the amplitude of the scalar field wave vector 
${\bf k}$. The energy-momentum dispersion relation is assumed of the form
\cite{viqar}:
\bean
\ep=\gamma\,\omega^\beta \label{emd}
\eean
where $\gamma$ and $\beta$ are constants to be 
determined. We now calculate the 
relevant thermodynamic quantities for the scalar field. It is well-known that
the mean occupation number $\overline{n_k}$ for a 
Bose gas is (see e.g. \cite{landau})
\bean
\overline{n_k}=\frac{1}{e^{(\ep_k-\mu)/T}-1}\, \label{mon}
\eean
where, $\ep_k$ is the energy in mode $k$, $\mu$ is the chemical potential, 
$T$ is the temperature of the gas, and we set the Boltzmann constant 
$k_{\rm B}=1$. 
Note that $\mu=0$ for the scalar field. Following a canonical ensemble 
standard calculation, we find that the free energy $F$ inside a volume $V$ is
\bean
F&=&\frac{TV}{2\pi^2}\int_0^\infty \omega^2 \log(1-e^{-
\gamma\omega^\beta/T})d\omega
\nonumber  \\
&=& -\frac{V \,T^{(3+\beta)/\beta}}{2\pi^2\,\beta\, \gamma^{3/\beta}}\,
\Gamma\left(\frac3\beta\right)\,\zeta\left(1+\frac3\beta\right)\,.
\label{fen}
\eean
Other thermodynamic quantities can be easily derived from the free energy 
$F$. First, we find the relations
\bean
P&=&\frac{\beta}{3}\,\rho \label{prh} \\
\rho &=&\frac{1}{36\,\gamma}\, T^2 \,. \label{eds}
\eean
From \eqs{prh} and \meq{per}, the constant $\beta$ can be identified
immediately as 
\bean
\beta=3\,. \label{bet}
\eean
Here, the constant $\gamma$ remains undetermined.  \eq{eds} is a
Stefan-Boltzmann law for a gas in one spatial dimension. In
\cite{viqar}, the coefficient of $T^2$ was indeed identified with a 
Stefan-Boltzmann constant $\sigma$, $\sigma=1/36\gamma$.

The entropy density, ${s_{}}_{\rm phys}$, takes the form 
\bean
{s_{}}_{\rm phys}=2\sigma T \,. \label{edes}
\eean
Combining \eqs{fov} , \meq{per}, \meq{eds} and \meq{edes}, we obtain
the entropy flux 4-vector: 
\bean
s^a={s_{}}_{\rm phys}\,u^a=-\sqrt{2\sigma}\,\partial^a \phi\,. \label{saf}
\eean
Husain \cite{viqar} chose $\sigma=\frac{1}{2}$ such that 
\bean
s^a=-\partial^a \phi. \label{enf}
\eean
In order to compare our results with that in \cite{viqar}, we shall 
follow this
choice and use \eq{enf} to compute entropy. 

\subsubsection{Testing entropy bounds for future-directed ingoing light 
sheets} \label{subfi}
We shall study the light sheets which  are generated by the
future-directed ingoing null vectors $k^a$ given in 
\eq{ig}. From  \eqs{phi}, \meq{ig} and \meq{enf}, we have the entropy density
associated with the null vector:
\bean
s=-s_ak^a=\frac{1}{4\sqrt{\pi}t^2} \left[\sqrt{3}\left(\frac{r-2}{r}\right)
^{\frac{\sqrt{3}}{2}}+\frac{t}{r^2-2r} \right]\,. \label{slm}
\eean
It is also straightforward to calculate $s'$ and $T_{ab}k^ak^b$.

\begin{figure}[bhtp]
\centering 
\scalebox{0.9}
{\includegraphics{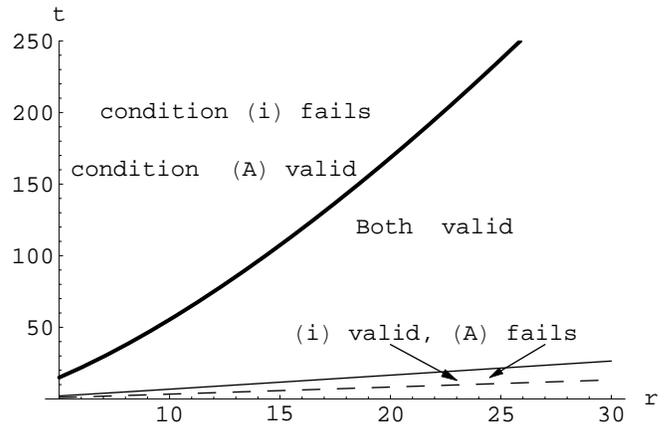}}
\caption{A spacetime diagram depicting the important regions 
under discussion.
Condition (i) is satisfied below the thick solid line, while 
condition (A) is valid above the thin 
solid line.  The dotted line represents the apparent horizon. 
} 
\label{ats}
\end{figure}
In \fig{ats} we show explicitly where these conditions are satisfied. 
We see that there exists a region where condition (A) is valid but
condition (i) fails. If a light sheet lies in this region, then the
covariant bound is satisfied, although condition (i) fails everywhere
on the light sheet. This illustrates that condition (i) in \cite{quan}  
is not a necessary condition for the entropy bound.

\begin{figure}
\centering
  \subfigure[]{ 
\scalebox{0.8}{ \includegraphics{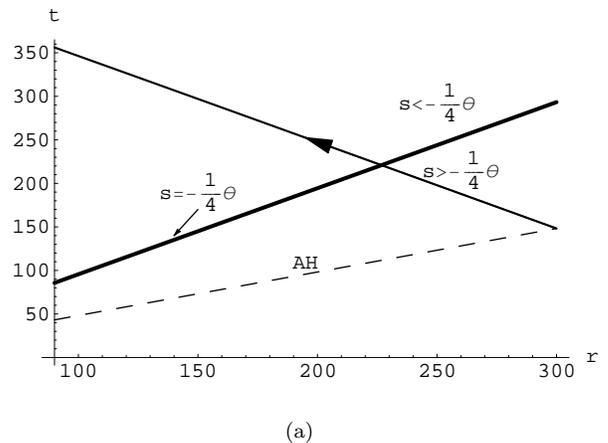}} } 
\hspace{2in}
\subfigure[]{
 \scalebox{0.8}{     \includegraphics{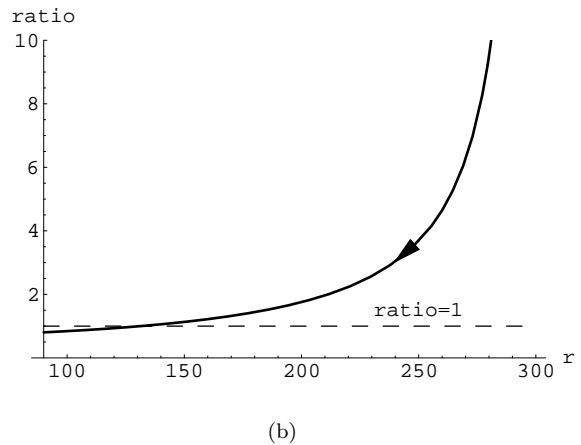}}}
\caption{(a) A light sheet crossing the $s=-\frac{1}{4}\theta$
  surface. AH represents the apparent horizon. (b) Change of ratio on
  the light sheet. 
 The generalized bound is violated in the
  $s>-\frac{1}{4}\theta$ zone and is satisfied until the light sheet
  runs certain distance in the $s<-\frac{1}{4}\theta$ zone. }
\label{croscr}
\end{figure}

An important point to note is that the Proposition \ref{prop} does not
cover the case when a light sheet crosses the $s=-\frac{1}{4}\theta$
line. In \fig{croscr}(a),  a light sheet starts near the apparent
horizon and crosses the $s=-\frac{1}{4}\theta$ line. This
standard light sheet has an initial 2-sphere with fixed radius $r_0$
and fixed area $A_0$. We can truncate this standard light sheet at any
other inner radius $r$ to get a truncated light sheet with a given
initial 2-sphere specified by $r_0$ and $A_0$.  
By proceeding inwards with this truncation we get a series of light sheets, 
each labeled by the coordinate $r$ and area $A(r)$ 
at which it is truncated, all starting at the same 
initial 2-sphere ($r_0$,$A_0$).
We shall test the bound for each one of them. Define
\bean
{\rm ratio}(r)=\frac{1}{4}\frac{S(r)}{A_0-A(r)}, \label{dfra}
\eean
where $S(r)$ is the entropy passing through the light sheet starting from the 
fixed 2-sphere with area $A_0$ and ending at the 2-sphere with area
$A(r)$. If ${\rm ratio}(r)\leq 1$, the generalized bound, \eq{get}, is 
satisfied. \fig{croscr}(b) plots the change of ${\rm ratio}\,(r)$. 
The $\rm ratio$ is larger than unity in the $s>-\frac{1}{4}\theta$ zone,
and it remains larger than unity after the light sheet enters the
$s<-\frac{1}{4}\theta$ zone until about $r=140$.

In order to decide whether this means a solid violation of the
generalized Bousso bound, we first check in which 
scale the fluid description of the matter fails. 
The entropy flux computation is invalid if the light sheet is shorter 
than the matter thermal wavelength \cite{viqar,simple}. 
The thermal wavelength $\lambda$ is estimated
from the relation $\lambda\sim 1/k$, 
where $k$ is the momentum of a fluid particle. From \eq{omk} 
and the energy-momentum
dispersion relation \eq{emd} ($\beta=3$), we have
\bean
\lambda\sim\frac{1}{\bar\epsilon^{1/3}} \label{leb}
\eean
where $\bar\epsilon$ is the mean energy per particle. To estimate
$\bar\epsilon$, we first calculate the particle number density, $n$, by
integrating the occupation number $\overline{n_k}$ (see \eq{mon}) over
all modes:
\bean
n&=&\frac{1}{\pi^2}\int_0^\infty \overline{n_k} \omega^2d\omega
\nonumber \\
&=&\frac{1}{\pi^2}\int_0^\infty \frac{1}{e^{\omega^3/T}-1} \omega^2
d\omega \nonumber \\
&\sim&T  \label{intn}
\eean
Noting that the energy density $\rho\sim T^2$ (see \eq{eds}), we have
immediately $\bar\epsilon=\rho/n\sim T$. Therefore, \eq{leb} gives 
the thermal wavelength $\lambda \sim T^{-1/3}$.
From \eq{eds}, we finally obtain 
\bean
\lambda \sim \rho^{-1/6}\,.
\label{lambdaofT}
\eean
(This is different from the length scale 
$\rho^{-1/4}$ given in \cite{viqar}, which follows 
a black body radiation calculation, but
we believe is not a consistent estimation for the scalar field.)

\begin{figure}
\centering
\scalebox{0.8}{  \includegraphics{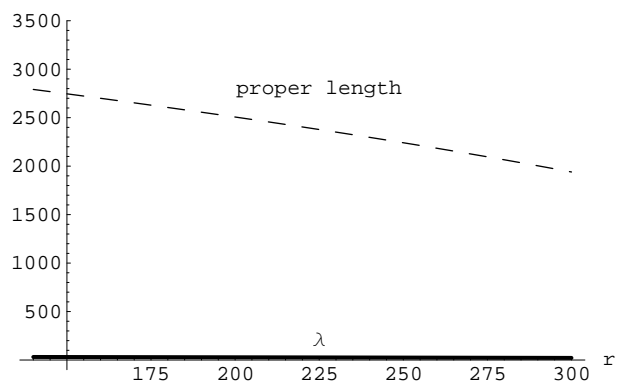}}
\caption{Plots of the proper length and thermal wavelength $\lambda$
  for the light sheet in \fig{croscr}(a). This figure shows that  
the proper length of the light sheet is significantly larger than the
thermal wavelength $\lambda$.}
\label{distem}
\end{figure}

Now, after this preamble, we consider the light sheet as in
\fig{croscr}(a). We let it start near the apparent horizon and
stop  at $r=140$ where the $\rm ratio$ is close to but larger 
than $1$ according to
\fig{croscr}(b). We define the proper length of the light sheet as the
length between two comoving observers, one that passes the starting
point (near the apparent horizon) and the other that passes the ending
point of the light sheet (near $r=140$). Since the comoving frame is
not static, this proper length, $L$, changes with time and can be easily
calculated from the metric \meq{st} as: 
\bean
L(t)=\int_{r_0}^{r_f}\sqrt{\frac{t}{f(r)}}, \label{lt}
\eean
where $r_0$ and $r_f$ are the radial coordinates of the two comoving
observers. Along the light sheet, $t$ is a function of $r$, so the
proper length, 
$L$, can be expressed as a function of $r$, as plotted in 
\fig{distem}. The thermal wavelength $\lambda$ is a function of the
temperature. The temperature at each point on the light-sheet is a
function of $r$, so $\lambda$ is a function of $r$. This is also plotted in
\fig{distem}.  Since $\lambda$ is always much smaller than the
proper length, the local entropy description is justified for the
light sheet. Thus the generalized Bousso bound is not valid in this case.

\subsubsection{Testing the entropy bounds for past-directed ingoing
  and past-directed outgoing light sheets}
We first investigate the generalized Bousso bound 
for the light sheets generated by past-directed ingoing
null geodesics. The tangent field of these null geodesics takes the
form:
\bean
k^a=-\frac{1}{f(r)\,t}\left(\pf{}{t}\right)^a-\frac{1}{t}
\left(\pf{}{r}\right)^a\,. \label{pastka}
\eean
The entropy density is $s=s_ak^a$ (note that the sign of this expression
is different from that in \eq{slm} for $k^a$ in \eq{pastka} is
past-directed), and the expansion for $k^a$ can be calculated
straightforwardly. To check condition (A), we consider the following
expression 
\bean
s+\frac{1}{4}\theta&=&\frac{1}{4(r-2)t^2}\left[
  \left(-1+\sqrt{\frac{3}{\pi}}\right)
    \left(\frac{r-2}{r}\right)^{1+\frac{\sqrt{3}}{2}} r
    \right. \nonumber \\ 
&+&\left. \left(2-2r-\sqrt
    3-\frac{1}{\sqrt{\pi}}\right)\frac{t}{r}\right] \,.
\eean
Since $r>2$, we see immediately that $s+\frac{1}{4}\theta<0$ in the
whole spacetime, i.e., condition (A) is satisfied
everywhere. Thus, according to Proposition \ref{prop}, the generalized
Bousso bound holds for all past-directed ingoing light
sheets. Consequently, the original Bousso bound holds for all
these light sheets. Therefore, with the help of Proposition
\ref{prop}, both the covariant entropy bounds for past-directed
ingoing light sheets have been easily tested. 

\begin{figure}[htpb]
\centering
\scalebox{0.8}
{\includegraphics{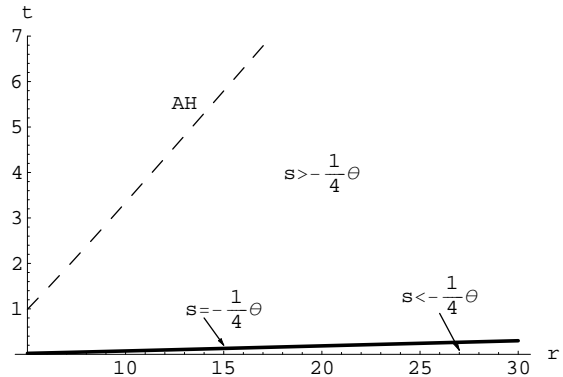}}
\caption{A spacetime diagram showing the region where condition (A)
  $s<-\frac{1}{4}\theta$ holds for past-directed outgoing light sheets.
}  \label{aps}
\end{figure}

Now we investigate the generalized Bousso bound 
for the light sheets generated by past-directed outgoing
null geodesics. This is much more complicated. 
The tangent field of these
null geodesics is 
\bean
k^a=-\frac{1}{f(r)\,t}\left(\pf{}{t}\right)^a+\frac{1}{t}
\left(\pf{}{r}\right)^a\,.
\eean
 These light sheets must be located in the past of the
apparent horizon and may terminate at the past singularity
$t=0$. \fig{aps} shows that condition (A) is violated in a
neighborhood of the apparent horizon.  Indeed, it holds only near the 
spacelike singularity. An analysis similar to the one done in section
\ref{subfi} shows that the generalized Bousso bound is violated for
light sheets between the apparent horizon and the
$s=-\frac{1}{4}\theta$ surface. One can also check that
condition (i) is nowhere satisfied in the past of the apparent
horizon.

\begin{figure}
\centering
 \subfigure[]{
\scalebox{0.8}{\includegraphics{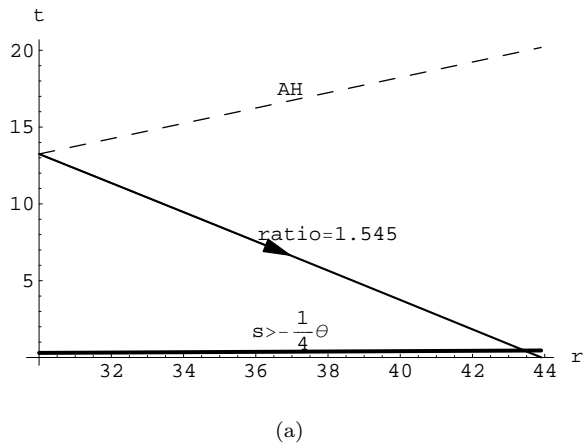}}}
\hspace{2in}
\subfigure[]{
 \scalebox{0.8}{\includegraphics{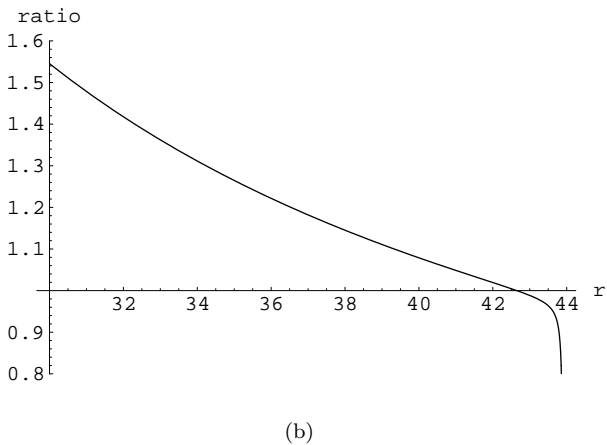}}}
\caption{(a) A past-outgoing light sheet located between the apparent
  horizon and the past singularity. (b) Plot of $\rm ratio(r)$ 
  when the light sheet in part (a) of this figure rolls down
  from the apparent horizon to the singularity. }
\label{tthpra}
\end{figure}

The importance and interest of past-directed (both ingoing 
and outgoing) light sheets is that it
enables us to investigate the original Bousso bound, since
these light sheets can reach the past singularity.  Consider then a
specific light sheet which starts at the apparent horizon with
coordinates $(t_0,r_0)=(13.25,30)$ and terminates at the past singularity with
coordinates $(t_f,r_f)=(0,43.9)$, see \fig{tthpra}(a). The $\rm ratio$ for
this light 
sheet is ratio$=1.545$, indicating that the covariant entropy bound
is violated, apriori.  
One can fix the ending 2-sphere at the singularity and
move continuously the initial 2-sphere along the 
original light sheet, labeling each initial 2-sphere by its coordinate
$r$, and test the Bousso bound for all these light sheets.  This is
done in \fig{tthpra}(b), where we obtain the  $\rm ratio(r)$ as a
function of the initial 2-sphere $r$. We see that the maximum
violation occurs when the light sheet starts at the apparent horizon,
the case shown in \fig{tthpra}(a).

\begin{figure}[htpb]
\centering
\scalebox{0.8}
{\includegraphics{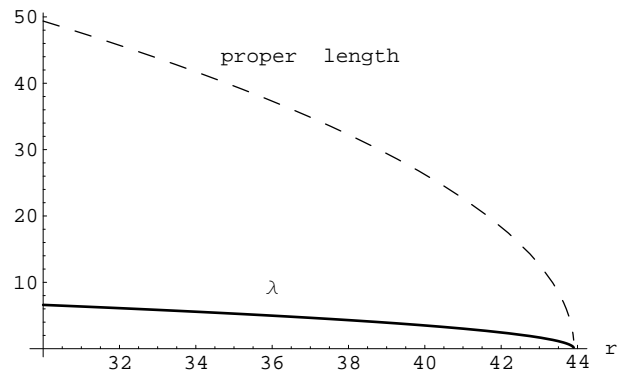}}
\caption{
Plots of the proper length and thermal wavelength $\lambda$
for the light sheet in \fig{tthpra}(a). We consider the light sheet as in
\fig{tthpra}(a), but we let it start near the apparent horizon and 
stop it at $r=43.9$ where the singularity is located.
Since the comoving frame is
not static, this proper length changes with time. But along the light
sheet, the time $t$ is a function of $r$. So the proper length can
be plotted as a function of $r$.  This figures shows that  
the proper length of the light sheet is significantly larger than the
thermal wavelength $\lambda$.
}  \label{pdt}
\end{figure}

Similarly to the discussion in section \ref{subfi}, we shall compare
the proper length of the light sheet with the matter thermal
wavelength. As plotted in \fig{pdt}, the proper length between the two
observers passing the light sheet changes with $r$.  As well the
matter thermal wavelength changes with $r$.  We see that the proper
length of the light sheet dominates $\lambda$ all the way.  Therefore,
the length scale argument does not save the covariant entropy bound.

We still have a trump on the sleeve, which is the Stefan-Boltzmann
constant $\sigma$. Remember we have followed Husain's choice $\sigma=1/2$ 
\cite{viqar}, but
this does not need to be the case. Since the entropy for any light
sheet is proportional to $\sqrt{\sigma}$ (see \eq{saf}), the bound could be
rescued by choosing a smaller value of   $\sigma$.  In Table
\ref{tab} we give the 
$\rm ratio$ for four light sheets that start at the apparent horizon
with coordinates $(t_0,r_0)$, and end at the spacelike singularity
with coordinates $(t,r)=(0,r_f)$, where $r=r_f$ is the radial
coordinate of the light sheets at the singularity.
It shows that the $\rm ratio$
for these (and by inference all) light sheets starting at the 
apparent horizon and ending at the
singularity is almost a constant. The reason behind this 
coincidence is unknown to us, but this fact is very instructive for
fixing the covariant bound. We can also let the initial 2-sphere of each
light sheet in Table \ref{tab} move away from the apparent horizon and  
plot the change of $\rm ratio(r)$ as a function of the initial 2-sphere 
labeled by $r$. It turns out that their behaviors are all
similar to that in \fig{tthpra}(b). Therefore, if the entropy density
\meq{saf} is scaled down by a factor 
smaller than $ 1/1.547$, the bound will be saved. This corresponds to
requiring $\sigma\leq 0.21$. Thus the covariant Bousso bound puts an
upper limit on the constant $\sigma$.  

\vskip 1cm
\begin{table}
\caption{\ \label{tab} The coordinates for the
  starting spherical surfaces at the apparent 
  horizon are $(t_0,r_0)$. Thus, all light sheets begin at the apparent
  horizon and end at the spacelike singularity 
  $(t,r)=(0,r_f)$, where $r=r_f$ is the radial 
  coordinate of the light sheets at the singularity.  }
\begin{ruledtabular}
\begin{tabular}{llll}
 $t_0$& $r_0$ & $r_f$ & $\rm ratio$ \\ \hline
 $13.25$&$30$  & $43.9$ & $1.545$ \\ 
 $1498$& $3000$ & $4498$ &$1.546$ \\ 
$5998$&$12000$  & $17998$ & $1.547$ \\
$59998$&$120000$   & $179998$ & $1.547$ \\ 
\end{tabular}
\end{ruledtabular}
\end{table}

\section{Conclusions}

We have investigated the sufficient conditions for the generalized
covariant entropy bound proposed by Strominger and Thompson \cite{quan}. 
We showed
that the condition $s(x)\leq -\frac{1}{4}\,\theta(x)$, 
our condition (A), can be used to identify
the regions where the generalized entropy bound is satisfied for all
light sheets. We applied this condition to a closed, dust-dominated
FRW universe and a scalar field spacetime. We have found that in the closed
FRW spacetime, condition (A) is satisfied in most of the
spacetime. Violations occur only in the regions very close to the
apparent horizon and the singularity.  According to our Proposition
\ref{prop}, the generalized Bousso bound is violated in these
regions. But such violations are due to the breakdown of the local
description of entropy and the breakdown of classical relativity.
Then, following the original investigation by Husain, we have studied the
covariant entropy bounds in a scalar field spacetime. Husain has found
that the generalized covariant entropy bound is violated only in a
band region around the apparent horizon surface. Proposition \ref{prop}
indicates that such a band region exists in spacetimes where the
matter entropy does not vanish in a neighborhood of the apparent
horizon. We also have checked the validity of the local description of entropy
for a light sheet which violates the generalized entropy bound. It
turns out that the proper length of the light sheet is much larger
than the thermal wavelength, meaning the entropy computation is valid
in this case. Our formula for the thermal
wavelength, $\lambda\sim\rho^{-1/6}$, is consistent with the dispersion
relation of the scalar field. This is different from Husain's
estimation $\lambda\sim\rho^{-1/4}$, which obviously follows a black
body argument. Husain showed that there is no
violation for the covariant 
entropy bound by calculating the entropy of light sheets hitting the
timelike singularity $r=2$. Our calculation shows that the comoving
observers are no longer timelike near the timelike
singularity. Consequently, there is no meaningful definition for
entropy in that region. In contrast, we have considered the past-directed
light sheets which hit the past singularity $t=0$. We have checked
condition (A) for past-directed ingoing light sheets and found it
holds for all of them. According to Proposition \ref{prop}, both the
generalized and the original Bousso bounds hold for such light
sheets. For past-directed outgoing light sheets starting near the
apparent horizon, violations of the bounds have been found. However,
these violations rely 
on the artificially selected Stefan-Boltzmann constant $\sigma$. Our
numerical results suggest that the covariant entropy bound can be
rescued by choosing a smaller value of $\sigma$. Therefore, the entropy
bound conjecture sets an upper bound on $\sigma$.

\begin{center}
{\bf  \large Acknowledgments}
\end{center}
This work was partially funded by Funda\c c\~ao para a Ci\^encia e
Tecnologia (FCT) - Portugal through project POCTI/FNU/44648/2002. 
SG acknowledges financial support by the FCT
grant SFRH/BPD/10078/2002 from FCT.   JPSL 
thanks Observat\'orio Nacional do Rio de Janeiro for hospitality.

\end{document}